\documentclass[10pt,fleqn]{article}

\usepackage{amssymb}

\newcommand{\name}[1]{\begin{flushleft}
                       \LARGE \bf #1
                       \end{flushleft}\vspace{-3mm}}

\newcommand{\Author}[1]{\begin{flushleft}
                       \it #1 \end{flushleft}}

\newcommand{\Adress}[1]{\begin{flushleft}
                       \it #1 \end{flushleft}}

\newcommand{\AbsEng}[1]{
    \begin{flushright}
    \begin{minipage}{120mm}
     \small   #1
    \end{minipage}
    \end{flushright}
}

\newcommand{\be}{\begin{equation}}
\newcommand{\ee}{\end{equation}}
\newcommand{\ba}{\hspace*{-5pt}\begin{array}}
\newcommand{\ea}{\end{array}}
\newcommand{\p}{\partial}
\newcommand{\ds}{\displaystyle}

\newcommand{\pbf}[1]{\mbox{\mathversion{bold}$#1$}}

\begin{document}

\name{On representations of the inhomogeneous\\
de Sitter group and equations in five-dimensional Minkowski space}

\medskip

\noindent{published in {\it Nuclear Physics B}, 1969, {\bf 14}, P. 573--585.}

\Author{Wilhelm I. FUSHCHYCH and Ivan Yu. KRIVSKY}

\Adress{Institute of Mathematics of the National Academy of
Sciences of Ukraine,\\ 3 Tereshchenkivska  Street, 01601 Kyiv-4,
UKRAINE}

\noindent {\tt URL:
http://www.imath.kiev.ua/\~{}appmath/wif.html\\ E-mail:
symmetry@imath.kiev.ua}

\AbsEng{This paper is a continuation and elaboration of our brief notice [1] where some
approach to the variable-mass problem was proposed. Here we have found a definite
realization of irreducible representations of the inhomogeneous group $P(1,n)$,
the group of translations and rotations in $(1+n)$-dimensional Minkowski space,
in two classes (when $P_0^2-P_k^2>0$ and $P_0^2-P_k^2<0$). All $P(1,n)$-invariant
equations of the Schr\"odinger--Foldy type are written down. Some equations of physical
interpretation of the quantal scheme based on the inhomogeneous de Sitter group $P(1,4)$
are discussed.

\hspace*{5mm}The analysis of the Dirac and Kemmer--Duffin type equations in the $P(1,4)$
sche\-me is carried out. A concrete realization of representations of the algebra $P(1,4)$
connected with this equations, is obtained. The transformations of the Foldy-Wouthuysen
type for this equations are found. It is shown that in the $P(1,4)$ scheme of the Kemmer--Duffin
type equation describes a fermion multiplet like the nucleon-antinucleon.}

\medskip

\centerline{\bf1. Introduction}

We recall here the initial points of our approach of the variable-mass problem proposed in ref.~[1]:

(i) The square of the variable-mass operator is defined as an independent dynamical variable:
\be
M^2 \equiv \varkappa^2 +P_4^2,
\ee
where $\varkappa$  is a fixed parameter and $P_4$ is an operator similar to the components
 of the three-momentum ${\pbf P}$, which commutes with all the generators of the algebra
$P(1,3)$ of the Poincar\'e group.

(ii) The relation between the energy $P_0$, three-momentum ${\pbf P}$
and variable-mass $M$ of a physical system remains conventional (here $\hbar =c=1$):
\be
P_0^2 ={\pbf P}^2+M^2 \equiv P^2 +\varkappa^2_k, \qquad k=1,2,3,4.
\ee

\renewcommand{\thefootnote}{\arabic{footnote}}
\setcounter{footnote}{0}

(iii) The spaces $p\equiv (p_0,p_1,\ldots, p_4)$ and $x=(x_0, x_1,\ldots, x_4)$
are assumed to be plane and reciprocally conjugated. It follows then from (i),
(ii) and (iii) that the generalized relativistic group symmetry is an inhomogeneous de Sitter
group\footnote{Algebras and groups connected with them are designated here with the
same symbols.} $P(1,4)$,
i.e. the group of translations and rotations in five-dimensional Minkowski space. This group
is a minimal extention of the conventional group of relativistic symmetry: the Poincar\'e group
$P(1,3)$.

In sect. 2 a definite realization of irreducible representations for the generators
$P_\mu$, $J_{\mu\nu}$ of the algebra $P(1,n)$ with arbitrary $n$  is carried out,
which made it possible to give a proof of the $P(1,n)$-invariance of the Schr\"odinger-Foldy
type equations given in ref.~[1] for $n=4$. Some questions of a physical interpretation of
a quantal scheme based on the group $P(1,4)$, are considered in sect.~3. Sects.~4 and~5 answer
the question which representations of the group $P(1,4)$ are realized by two types of
equations linear in $\p_\mu \equiv \p/\p x_\mu$ --- the Dirac and Kemmer--Duffin type equations.

\medskip

\centerline{\bf 2. Realizations of the algebra $\pbf{P(1,n)}$ representations}

For the sake of generality all considerations are made here not for the de Sitter group
$P(1,4)$ but for the group $P(1,n)$ of translations and rotations in $(1+n)$-dimensional
Minkowski space which leaves the form
\be
\ba{l}
x^2 \equiv x_0^2-x_1^2-\cdots -x_n^2 \equiv x_0^2-x_k^2 \equiv x_\mu^2,
\vspace{1mm}\\
\mu =0,1,2,\ldots, n, \qquad k=1,2,\ldots, n,
\ea
\ee
unchanged, where $x_\mu$ are differences of point coordinates of this space.

Commutation relations for the generators $P_\mu$, $J_{\mu\nu}$ of the algebra
$P(1,n)$ are choosen in the form
\renewcommand{\theequation}{\arabic{equation}{\rm a}}
\setcounter{equation}{3}
\be
\left[ P_\mu, P_\nu\right]=0, \qquad
-i\left[ P_\mu, J_{\rho\sigma}\right]=g_{\mu\rho} P_\sigma-g_{\mu\sigma}P_\rho,
\ee
\renewcommand{\theequation}{\arabic{equation}{\rm b}}
\setcounter{equation}{3}
\be
-i \left[ J_{\mu\nu}, J_{\rho\sigma}\right] =g_{\mu\sigma} J_{\nu\rho} +g_{\nu\rho}J_{\mu\sigma}
-g_{\mu\rho}J_{\nu\sigma} -g_{\nu\sigma} J_{\mu\rho},
\ee
where $g_{00}=1$, $-g_{kl}=\delta_{kl}$, $P_\mu$ are operators of infinitesinal
displacements and $J_{\mu\nu}$ are operators of infinitesimal rotations.

In refs. [2--5] all irreducible representations of the Poincar\'e group $P(1,3)$
are studied and the concrete realization for the generators of its algebra is found.
The methods are generalized here for the case of the group $P(1,n)$.

For representations of the class I $(P^2\equiv P_0^2 -P_k^2>0)$ when the group $O(n)$
of rotations in $n$-dimensional Euclidean space is the little group of the group $P(1,n)$,
the generators $P_\mu$, $J_{\mu\nu}$ are of the form
\renewcommand{\theequation}{\arabic{equation}}
\setcounter{equation}{4}
\be
\ba{l}
P_0=p_0 \equiv \varepsilon \sqrt{p_k^2-\varkappa^2}, \qquad P_k=p_k,
\vspace{1mm}\\
J_{kl} =x_{[k} p_{l]}+S_{kl}, \qquad x_{[k}p_{l]}\equiv x_k p_l-x_l p_k,
\vspace{1mm}\\
\ds J_{0k}=x_0 p_k -\frac 12 (x_k p_0 +p_0 x_k) -\frac{S_{kl} p_l}{p_0 +\varkappa},
\ea
\ee
where the operators $x_k$ and $p_k$ are defined by the relations
\be
[x_k, p_l]=i\delta_{kl}, \qquad [x_k, x_l]=[p_k, p_l]=0,
\ee
and $S_{kl}$ are matrices realizing irreducible representations of the algebra $O(n)$
which have been studied in ref.~[6].

For representations of the class III $(P^2=P_0^2-P_k^2<0)$ when the little group of the group
$P(1,n)$ is already a non-compact group $O(1,n-1)$ of rotations in $[1+(n-1)]$-dimensional
pseudo Euclidean space, the generators $P_\mu$, $J_{\mu\nu}$ are of the form
\be
\ba{l}
\ds P_0=p_0 \equiv \pm \sqrt{p_k^2-\eta^2}, \qquad P_k=p_k,
\qquad
J_{ab} =x_{[a}p_{b]} +S_{ab},
\vspace{1mm}\\
\ds J_{an} =x_{[a}p_{n]} -\frac{S_{ab} p_b -S_{a0}p_0}{p_n +\eta},
\qquad
\ds J_{0a}=x_0 p_a -\frac 12 (x_a p_0 +p_0 x_a) +S_{0a},
\vspace{1mm}\\
\ds J_{0n} =x_0 p_n -\frac 12 (x_n p_0 +p_0 x_n) -\frac {S_{0a} p_a}{p_n +\eta},
\ea
\ee
where $a,b=1, \ldots, n-1$, $\eta$  is a real constant, the operators $x_k$, $p_k$
are defined by relations~(6) as before, and the operators $(S_{0a}, S_{ab})$ are generators
of the algebra $O(1,n-1)$ in corresponding irreducible representations, which have been
studied by Gelfand and Grayev~[7].

Formulae (5) and (7) give the irreducible representations of the algebra  $P(1,n)$  in the
Schr\"odinger picture: a representation space for an irreducible representation is
constituted from the solutions  $\Psi(x_0)$    of the Schr\"odinger--Foldy type equation
\be
i\p_0 \Psi(x_0) =P_0 \Psi(x_0).
\ee
The solutions $\Psi(x_0)$ are vector functions $\Psi(x_0)= \Psi(x_0, x_1,\ldots, x_n)$
in $x$-representation for eq.~(6) or $\Psi(x_0) =\widetilde \Psi (x_0, p_1,\ldots, p_n)$ in
$p$-represen\-ta\-tion for eq.~(6) etc., and their components are also functions of auxiliary
variables $s_3, t_3,\ldots$  (``spin'' variables) --- eigenvalues of generators of Cartan's
subalgebra of the algebra $O(n)$ in the case~(5) or $O(1,n-1)$ in the case~(7).

Eq. (8) is $P(1,n)$-invariant: the manifold of all the solutions of eq.~(8) is invariant under
transformations from the group $P(1,n)$. This is the consequence of the condition
\be
[(i\p_0 -P_0), Q]\Psi =0
\ee
being valid for any generator $Q$ of $P(1,n)$ defined by eqs. (5) or (7).

In the Heisenberg picture where vector functions $\Psi$ of a representation space for a
representation of $P(1,n)$ do not depend on the time $x_0$ (and are the solutions of
the equation $P_0 \Psi =E\Psi$), formulae for the generators $P_\mu$, $J_{\mu\nu}$
are obtained by dropping the terms with $x_0$, and eq.~(8) is replaced by
\be
i\p_0 Q= [Q, P_0]_-
\ee
for any operators $Q$ as functions of $x_k$, $p_k$, $S$.

Since in class I the little group of the group $P(1,n)$ is the compact group
$O(n)$, all the irreducible representations of the group $P(1,n)$ are here unitary and
finite-dimensional (concerning a set of ``spin'' indexes $s_3,t_3,\ldots$),
and the solutions of the corresponding equation~(8) have here finite number of components.
In accord with the representations of the little group $O(1,n-1)$, in class III the group
$P(1,n)$ has both finite- and infinite-dimensional representations. We emphasize that all
the unitary representations are here infinite-dimensional, and the solutions of the
 corresponding eq.~(8) have here infinite number of components.

Note at the end of this section that the problems of classification and realization of
representations of an arbitrary inhomogeneous group $P(m,n)$ can similarly, without principal
difficulties, be reduced to problems of classification and realization of cor\-re\-spon\-ding
representations of homogeneous group of the types $O(m', n')$.

\medskip

\centerline{\bf 3. Physical interpretation}

Here we deal only with the inhomogeneous de Sitter group $P(1, 4)$ which is a minimal extention
of the Poincar\'e group $P(1,3)$. We discuss the main topics of the physical interpretation
of a quantal scheme based on the group $P(1,4)$. This group is the most attractive one
because it will succeed to give a clear physical meaning to a complete set of commuting
variables.

ln the $p$-representation for eq.~(6) a component of the wave function $\Psi$ ---
a solution of eq.~(9) with $n=4$ --- is a function of six dynamical variables of the
corresponding complete set: $\Psi(x_0,{\pbf p}, p_4, s_3,t_3)$ where ${\pbf p}$
and $p_4$ are eigenvalues of the operators ${\pbf P}$ and $P_4$ and their physical
meaning has been discussed in the introduction; $s_3$ and $t_3$ are eigenvalues of the third
components of the operators ${\pbf S}=(S_1, S_2, S_3)$ and ${\pbf T}=(T_1, T_2, T_3)$
where
\be
S_a \equiv \frac 12 (S_{bc} +S_{4a}), \qquad T_a\equiv \frac 12 (S_{bc} -S_{4a}),
\ee
$(a,b,c)=\mbox{cycl}(1,2,3)$.  These operators satisfy the relations
\be
[S_a, S_b]=iS_c, \qquad [T_a,T_b]=iT_c, \qquad
[S_a, {\pbf S}^2] =[T_a, {\pbf T}^2]=[S_a, T_b]=0.
\ee
The operators
\be
{\pbf S}^2 =\frac{W}{4p^2} +\frac{V}{2\sqrt{p^2}}, \qquad
{\pbf T}^2 =\frac{W}{4p^2} -\frac{V}{2\sqrt{p^2}}
\ee
are invariant both of $P(1,4)$ and $O(4)$  (in class I) or $O(1,3)$ (in class III). Note that
in irreducible representations of class I we have
\be
{\pbf S}^2 =s(s+1)\hat 1, \qquad {\pbf T}^2 =t(t+1)\hat 1,
\ee
where $s,t=0,\frac 12, 1,\ldots,\ldots$ and $\hat 1$  is the $(2s+1)(2t+1)$-dimensional unit matrix.

The irreducible representations $D^\pm (s,t,\varkappa^2)$ of the group $P(1,4)$,
identified by fixed numbers $s$, $t$, $\varkappa^2$ and $\varepsilon =\pm 1$
(i.e., by values of the corresponding invariants of $P(1,4)$),
allow us to introduce the concept of ``elementary particle'' in the quantum scheme based on
the group $P(1,4)$ possible states of an ``elementary particle'' (when $\varepsilon =+1$)
or ``antiparticle'' (when $\varepsilon =-1$) with given values of $s$, $t$ and $\varkappa^2$,
are states which constitute the representation space for the irreducible representation
$D^\pm(s,t,\varkappa^2)$ of the group $P(1,4)$.
As it is seen from eq.~(2), $\varkappa$ is the boundary value of the energy $P_0$;
the physical meaning of $s$ and $t$ is dictated by the relations~(12): they allow to interpret
the operators ${\pbf S}$ and ${\pbf T}$ as the spin and isospin operators. Thus,
the components $\psi(x_0, {\pbf p}, p_4, s_3,t_3)$ are interpreted as the probability amplitude
of finding (by measuring at a given instant $x_0$) the indicated values of three-momentum
${\pbf p}$, mass $m=\sqrt{p_4^2+\varkappa^2}$ and third components of spin $s_3$
 and isospin $t_3$.

It is clear that an irreducible representation $D^\pm (s,t,\varkappa^2)$ of the group $P(1,4)$
is reducible with respect to $P(1,3)\subset P(1,4)$; therefore the ``elementary particle''
defined here, is not elementary in the conventional sense (i.e., with respect to the group
$P(1,3)$). The vector function of the representation space for $D^\pm(s,t,\varkappa^2)$
describes, in fact, a multiplet of particles with different $t_3$, $-t\leq t_3\leq t$
(and, of course, with different $s_3$, $-s\leq s_3\leq s$); the parameter  $\varkappa$
is then a ``bare'' rest mass of the given multiplet.

The $P(1,4)$ quantum scheme in our interpretation may be found successful for a
consequent description of unstable systems (resonances, particles or systems with
non-fixed mass) already in the framework of the quantal
approach\footnote{The consequent consideration of such problems demands, obviously,
the quantum field approach, but a quantal approach can be regarded as half-phenomenological.}
without breaking down such fundamental principles as unitarity, hermiticity etc. Indeed, here
the mass operator is an independent dynamical variable eq.~(1), it is Hermitian, and
the problem of unstable systems is, in fact, reduced to the problem of calculation quantities
like distributions
\be
\rho(m^2,s_3,t_3) \equiv \int d^3 x \left| \int d x_4 e^{-i\sqrt{m^2-\varkappa^2}x_4}
\Psi(x_0,\ldots, x_4,s_3,t_3)\right|^2,
\ee
where $\psi$ are solutions of an equation of the type~(8) with a suitable interaction. The
positions and forms of maxima of the distribution $\rho(m^2)$ define experimentally observed
masses and lifetimes of unstable particles, and singularities of $\rho(m^2)$
define masses of stable particles.

It is important to emphasize that in accord with our interpretation, the particles experimentally
observed are described not by the free equation~(8), but by an equation of the type~(8) with a
suitable interaction which may break the $P(1,4)$-invariance, but, of course, conserves the
$P(1,3)$-invariance\footnote{In this sense the consideration of $P(1 ,4)$ symmetry here
presented is only a base for its suitable violation --- analogously to considerations and violations
of $SU(n)$ symmetries.}. As for solutions of the free equation~(8), they are some
hypothetical (``bare'') states which may not correspond to any real particles. From the
viewpoint of this interpretation there are two types of interactions: interactions which cause
a ``dressing'' of particles and are inherent even in asymptotical states, and usual
interactions which cause a scattering processes of real (``dressed'') particles. Therefore,
in particular, the five-dimensional conservation law following from the free $P(1,4)$-invariant
scheme, may have no real sense.

We emphasize that the interpretation of the $P(1,4)$-scheme proposed does not pretend to
be the only one and complete. The more detailed discussions of in\-ter\-pre\-ta\-tion problems
are possible only in connection with solutions of suitable models of interactions in this
scheme, what is not the subject of this article.

\medskip

\centerline{\bf 4. The Dirac-type equations}

A characteristic feature of eqs. (8) is that they do not contain any redundant components.
However, in this equation the differential operators $\p_k\equiv \p/\p x_k$ enter under the
square root, therefore they are considered not to be appropriate for in\-tro\-du\-cing interactions
and for theoretical field considerations. Let us consider the simplest equation of first order in
$\p_\mu$, manifestly invariant under the group $P(1,4)$.

Remind that there are five Dirac matrices $\gamma_\mu$ satisfying the relations
\be
\gamma_\mu \gamma_\nu +\gamma_\nu \gamma_\mu =2 g_{\mu\nu}, \qquad
\mu,\nu =0,1,\ldots, 4,
\ee
where
\be
\gamma_0 \equiv \beta =\gamma_1\gamma_2 \gamma_3 \gamma_4
\qquad \mbox{or} \qquad \gamma_4 =-\gamma_0\gamma_1\gamma_2\gamma_3.
\ee
The Dirac equation in the Minkowski five-space is of the form
\renewcommand{\theequation}{\arabic{equation}{\rm a}}
\setcounter{equation}{17}
\be
(i\gamma_\mu \delta_\mu -\varkappa)\psi \equiv (i\gamma_0 \p_0 +i\gamma_k \p_k -
\varkappa)\psi =0
\ee
or
\renewcommand{\theequation}{\arabic{equation}{\rm b}}
\setcounter{equation}{17}
\be
(i\gamma_\mu \p_\mu +\varkappa)\psi =0.
\ee

Eqs. (18) were written down long ago by Dirac~[8]. It is clear that they are invariant under
the inhomogeneous de Sitter group. Our aim is to find out which representation of the group
$P(1,4)$ is realized in the representation space of solutions of eq.~(18). Here we shall not
follow the conventional method which is ordinary used (see, e.g.. refs.~[9,~10]) and which
in fact answers only the question which representation of the homogeneous Lorentz group
$O(1,3)$ is realized by the Dirac equation in the Minkowski four-space but does not answer
the question of representation of the Poincar\'e group $P(1,3)$.

Here we deal with the method suitable for analysis both of the Dirac equation and of other
wave equations (linear and non-linear with respect to $\p_\mu$) and besides in arbitrary
$(1+n)$-dimensional Minkowski space. The method is based on definition~(9) of the
invariance of the wave equation. It is clear from this definition that to answer the
question whether a wave equation is invariant under the group $P(1,n)$,
one has to find an explicit form of generators $P_\mu$, $J_{\mu\nu}$ of the algebra
connected with the equation in such a way that its Hamiltonian $H$  and the operator
$i\p_o \equiv i\p/ \p t$ must commute with the generators $P_k$, $J_{\mu\nu}$
 exactly just as the generator $P_0$  does. Further, if the explicit form of the generators
are found, one can find the invariants of the group $P(1,n)$ in the explicit form; their
eigenvalues will answer the question which representation of the group is realized by
solutions of this equation.

Let us illustrate the method for the case of eq. (18a). Rewrite eq.~(18a) in the Hamiltonian form
\renewcommand{\theequation}{\arabic{equation}{\rm a}${}'$}
\setcounter{equation}{17}
\be
i\p_0 \psi =H\psi , \qquad H\equiv \alpha_k p_k +\beta \varkappa, \qquad \alpha_k=\beta
\gamma_k.
\ee
It can, be immediately verified that in this case the explicit form of generators
$P_\mu$, $J_{\mu\nu}$  satisfying the relations of the algebra $P(1,4)$, is given by
\renewcommand{\theequation}{\arabic{equation}}
\setcounter{equation}{18}
\be
\ba{l}
P_0 =H \equiv \alpha_k p_k +\beta \varkappa, \qquad P_k =p_k \equiv -i\p_k,
\vspace{1mm}\\
J_{kl} =x_{[k} p_{l]} +S_{kl},
\qquad \ds J_{0k} =x_0 p_k -\frac 12 (x_k P_0 +P_0 x_k),
\ea
\ee
where
\be
S_{kl}\equiv \frac i4 (\gamma_k \gamma_l -\gamma_l \gamma_k).
\ee

We choose $\gamma_\mu$ in the form
\be
\gamma_a=\left( \begin{array}{cc}
0 & \sigma_a \\
-\sigma_a & 0 \end{array}\right), \qquad
\gamma_4 =i \left( \begin{array}{cc}
0 & 1\\
1 & 0
\end{array} \right), \qquad \gamma_0 \equiv \beta =\left(\begin{array}{cc}
1 & 0\\
0 & -1
\end{array}\right).
\ee
Then the spin and isospin operators for the particle described by eqs.~(18), are of the form
\be
S_a\equiv \frac 12 (S_{bc}+S_{4a})=\frac 12 \left( \begin{array}{cc}
\sigma_a & 0\\
0 & 0
\end{array} \right), \qquad \!
T_a\equiv \frac 12 (S_{bc}-S_{4a})=\frac 12 \left( \begin{array}{cc}
0 & 0\\
0 & \sigma_a
\end{array} \right),
\ee
and their squares coinciding with the invariants of the group $P(1,4)$,
are of the form
\be
{\pbf S}^ 2=\frac 34 \left( \begin{array}{cc}
1 & 0\\
0 & 0
\end{array} \right), \qquad
{\pbf T}^2=\frac 34 \left( \begin{array}{cc}
0 & 0\\
0 & 1
\end{array} \right).
\ee
Further, the invariant $P^2=\varkappa^2$ and the invariant $\varepsilon$ is the sign
of energy coincides (in the ``rest frame'' $p_k=0$) with the matrix $\beta$.

It is clearly seen from ${\pbf S}^2$, ${\pbf T}^2$, $S_3$, $T_3$ and $\varepsilon=\beta$
that the manyfold of solutions of eq.~(18a) constitutes the representation space for the
four-dimensional reducible representation $D^+(\frac 12,0)\oplus D^-(0,\frac 12)$
of the group $P(1,4)$. Thus, in accord with our interpretation of the numbers $s$ and $t$
the Dirac equation~(18a) describes a multiplet
\be
\psi =\left( \begin{array}{c}
\psi^+_{\frac 12, 0}
\vspace{1mm}\\
\psi^-_{0,\frac 12}
\end{array} \right),
\ee
where $\psi^+_{\frac 12, 0} $ is a spinor-isoscalar describing a fermion with the spin $s=\frac 12$
and isospin $t=0$ (a particle like the $\Lambda$ hyperon) and $\psi^-_{0,\frac 12}$
is a scalar-isospinor describing an antiboson with $s=0$ and $t=\frac 12$
(an antiparticle like the $\overline K$ meson)\footnote{Note that it would be more appropriate
to call the boson-like $K$ a spinosinglet-isodoublet, and the fermion-like $\Lambda$
a spinodoublet-isosinglet.}.

It can analogously be shown that eq. (18b) realizes the representation
$D^-(\frac 12,0)\oplus D^+(0,\frac 12)$ of the group $P(1,4)$,
i.e., describes a multiplet like $(K, \widetilde \Lambda)$. In this case the explicit form of the
operators $P_\mu$, $J_{\mu\nu}$
is obtained from eq.~(19) by the replacement $\varkappa\to-\varkappa$ or
$\beta\to -\beta$.

Thus, in contrast to the Dirac equations in the $P(1,3)$ scheme, the Dirac equations (18) in the
$P(1,4)$ scheme do not describe particles and antiparticles symmetrically and therefore
they will not be invariant under transformations of type $PTC$.

It can be perceived from the analysis of eqs. (18a) and (18b) that in the $P(1, 4)$
 scheme the equation describing particles and antiparticles symmetrically, must realize the
representation
\be
D^+\left( \frac 12,0\right) \oplus D^+\left(0, \frac 12 \right) \oplus D^-\left(\frac 12, 0\right)
\oplus D^-\left( 0,\frac 12\right).
\ee
We have found that such an equation is of the form
\be
(i\Gamma_\mu \p_\mu -\varkappa)\Psi \equiv (i\Gamma_0 \p_0+i \Gamma_k \p_k -
\varkappa) \Psi =0,
\ee
where the $8\times 8$ matrices are
\be
\Gamma_k =\left( \begin{array}{cc}
0 & \gamma_k\\
\gamma_k & 0
\end{array}
\right), \qquad \Gamma_0 =\left( \begin{array}{cc}
1& 0\\
0 & -1
\end{array}\right).
\ee

In the case of eq. (26) the explicit form of the generators of $P(1,4)$
is obtained from eq.~(19) by the replacement $\gamma_\mu \to \Gamma_\mu$.
One can see from the explicit form the the $8\times 8$ matrices
${\pbf S}^2$, ${\pbf T}^2$, $S_3$, $T_3$ and $\varepsilon=\Gamma_0$
that eq.~(26) actually realizes the representation~(25), i.e., that the wave function $\Psi$
(eight-component spinor) has the form
\be
\Psi =\left( \begin{array}{c}
\psi^+_{\frac 12, 0}
\vspace{1mm}\\
\psi^+_{0,\frac 12}
\vspace{1mm}\\
\psi^-_{0,\frac 12}
\vspace{1mm}\\
\psi^-_{\frac 12,0}
\end{array} \right),
\ee

Note, that in the $P(1,4)$ scheme just the eight-component equation~(26) (but not the
four-component equations~(18)) symmetrically describes particles and antipar\-tic\-les
and is therefore $PTC$ invariant (more detailed see refs.~[11,~16]).

It is easy to see that the eight-component equation (26) is the unification of the
four-component equations~(18a) and~(18b). Of course, in the $P(1,3)$
scheme such a unification of the Dirac equations is trivial. However, in the $P(1,4)$
scheme the unification is not trivial: the matrices $\Gamma_0$, $\Gamma_k$
obey the relations~(16), but they are not matrices of a reducible representation of the
Dirac algebra eq.~(16) since, in particular, $\Gamma_0\not= \Gamma_1 \Gamma_2 \Gamma_3
\Gamma_4$, i.e., the condition~(17) is not satisfied. The $8\times 8$ matrices
$\Gamma_0$, $\Gamma_k$ together with the two other matrices
\be
\Gamma_5 =i\left( \begin{array}{cc}
0 & \gamma_0\\
\gamma_0 & 0
\end{array}
\right), \qquad \Gamma_6 =\left( \begin{array}{cc}
0& 1\\
-1 & 0
\end{array}\right)
\ee
obey the commutation relations of Clifford algebra in six-dimensional space,
the additional condition
\renewcommand{\theequation}{\arabic{equation}${}'$}
\setcounter{equation}{16}
\be
\Gamma =-i\Gamma_1 \Gamma_2 \Gamma_3 \Gamma_4 \Gamma_5 \Gamma_6
\ee
being valid, and realize its irreducible representation. It is of interest to note that the
eight-component equation of the Dirac type
\renewcommand{\theequation}{\arabic{equation}}
\setcounter{equation}{29}
\be
(i\Gamma_\mu \p_\mu -\varkappa) \Phi \equiv (i \Gamma_0 \p_0 +
i\Gamma_1 \p_1 +\cdots + i \Gamma_6 \p_6 -\varkappa)\Phi=0
\ee
realizes a representation of the group $P(1, 6)$.

The wave function of eq. (26) (or even eqs. (18)) describes an unusual multiplet:
it unificates fermions and bosons into a multiplet. For example,
\be
\Psi=\left( \begin{array}{c}
\Lambda\\
K\\ \overline  \Lambda \\ \overline K
\end{array}\right).
\ee
This circumstance is not unsatisfactory for eq. (26) from the viewpoint of, for example,
the barion number conservation law. The latter only causes some restrictions on
possible forms of interactions. In the $P(1,4)$ scheme the barion number operator can be
defined as usually (as a number of fermions $\psi^+_{\frac 12,0}$ minus a number of
antifermions $\psi^-_{\frac 12,0}$). It is remarkable that the wave function~(28) describes
symmetrically both fermions and isofermions. Therefore in the $P(1,4)$ scheme we can
naturally define the operator of hypercharge as a number of isofermions
$\psi^+_{0,\frac 12}$ minus a number of anti-isofermions $\psi^-_{0,\frac 12}$.
This allows eq.~(26) to be considered as a fundamental equation for the dynamical approach
to the classification scheme of d'Espagnat and Prentki~[12].

As in the case of the Dirac equation in the $P(1,3)$ scheme~[13], in order to give an adequate
physical interpretation of the wave function $\Psi$  as a function of ${\pbf x}$, $x_4$,
one has to transit from the Dirac representation to the Foldy representation. The transition
is performed by the unitary transformation
\be
U=\exp\left( -i\frac{A_k p_k}{2p} \; \mbox{arctg}\; \frac{p}{\varkappa}\right),
\qquad p= \sqrt{p^2_k},
\ee
where $A_k =i\gamma_k$ for eqs.~(18) and $A_k =i\Gamma_k$ for eq.~(26).

In the Foldy--Shirokov representation eqs. (18) and~(26) are of the form
\be
i\p_0 \Psi =B\sqrt{p_k^2+\varkappa^2}\Psi,
\ee
where  $B=\gamma_0, -\gamma_0, \Gamma_0$ for eqs. (18a), (18b) and (26) correspondingly.

After the transformation (32), the formulae for the generators $P_\mu$, $J_{\mu\nu}$
coincide with eq.~(5) for $n=4$, if the replacement $\varepsilon \to B$ is made there.

\medskip

\centerline{\bf 5. The Kemmer--Duffin type equations}

Let us consider now an analogue of equations describing bosons with spin 0 and 1 in the
$P(1,3)$ scheme, namely, the equations in Minkowski five-space which are of the form
\be
(\beta_\mu \p_\mu +\varkappa)\Phi =0, \qquad \mu =1,2,3,4,5,
\ee
where five Hermitian matrices $\beta_\mu$ obey the algebra of the Kemmer-Duffin-Petiau type
(KDP):
\be
\beta_\mu \beta_\nu \beta_\lambda+\beta_\lambda \beta_\nu \beta_\mu=
\delta_{\mu\nu} \beta_\lambda +\delta_{\lambda \nu} \beta_\mu.
\ee

This algebra has three irreducible representations. The lowest representation is realized by
$6\times 6$ matrices. The non-zero element of these matrices are schematically written
down in table~1 where, for example, ``1,6'' denotes $(\beta_1)_{1,6}=1$.
Remind for comparison that the lowest representation of KDP algebra in the $P(1, 3)$
scheme (i.e., when $\mu\leq 4$) is realized by $5\times 5$ matrices.

\begin{center}
\small
{Table 1}\\
{The unit elements of $6\times 6$ matrices}\\[2mm]
\begin{tabular}{ccccc}
\hline
& &&&\\[-2mm]
$\ \beta_1\ $ & $\ \beta_2\ $ & $\ \beta_3\ $ & $\ \beta_4\ $ & $\ \beta_5\ $\\[1mm]
\hline
&&&&\\[-2mm]
1,6& 2,6 & 3,6 & 4,6 & 5,6  \\
6,1 & 6,2 & 6,3& 6,4& 6,5\\[1mm]
\hline
\end{tabular}
\end{center}

It can be shown by means of the method used in sect. 4 that eq.~(34) with the $6\times 6$
matrices~(35) realizes the representation
\be
D^+(0,0) \oplus D^-(0,0) \oplus D\left(\frac 12,\frac 12\right),
\ee
where the first two representations are realized by principal components of the vector function
$\Phi$, on which the energy operator has non-vanishing eigenvalues, and the last
representation is realised by redundant components of the vector function $\Phi$,
 on which the eigenvalues of the energy operator are equal to zero.
Thest last have no physical sense but they are presented in all linear, with respect to $\p_\mu$,
equations, except for the Dirac-type equations. In such cases the Foldy--Wouthuysen
transformation does not only split the states with positive and negative energies, but also
makes it possible to omit the redundant components by an invariant way.

Thus, eq. (34) with the $6\times 6$ matrices (35), which can be obtained by a linearization
procedure of the Klein--Gordon equation in the Minkowski five-space for a scalar, describes
particles and antiparticles with $s=t=0$.

Consider now a very interesting case: eq. (34) with the $15\times 15$ matrices $\beta_\mu$
realizing an irreducible representation of algebra~(35). These matrices can be taken,
for exam\-p\-le, in the form schematically given by table~2 where only the non-zero elements of
the matrices $\beta_\mu$ equal to $\pm 1$, are written down.

Now we shall find out what representation of the groups $O(4)$ and $P(1,4)$
is realized by solutions $\Phi$  of eq.~(34) with matrices $\beta_\mu$ of table~2.

Using the method proposed in ref. [14] for reducing the Kemmer-Duffin equations in the
$P(1,3)$ scheme to the Schr\"odinger form, one can verify that eq.~(34) is equivalent to the
equation
\be
i \p_0 \Phi =H\Phi, \qquad H=S_{5k}p_k +\beta_5 \varkappa,
\ee
where
\[
S_{5k} \equiv i (\beta_5 \beta_k -\beta_k \beta_5), \qquad k=1,2,3,4.
\]

Using eq.~(35) it is easy to verify that the Hermitian matrices
\be
S_{\mu\nu} \equiv i(\beta_\mu\beta_\nu-\beta_\nu \beta_\mu), \qquad  \mu,\nu =1,2,3,4,5,
\ee
obey the commutation relations for the generators of algebra, i.e., they realize a
fifteen-dimensional representation of this algebra. The explicit form of the matrices
$\beta_\mu$ given by table~2 allows to find the quantity corresponding to the invariant
$P^2$  of $P(1, 4)$
\be
P^2 \equiv H^2 -p_k^2 =\varkappa^2 \beta_5^2 ,
\qquad \beta_5^2 =\left(
\begin{array}{cccc}
1^4 & &&\\
& 0^6 & &\\
&& 1^4 & \\
&&& 0^1
\end{array} \right),
\ee
where the upper indexes denote the dimensionality of the matrices. Table~2 allows us also
to find the matrices $S_{kl}$.

\begin{center}
\small
\begin{tabular}{cccccccccc}
\multicolumn{10}{c}{Table 2}\\
\multicolumn{10}{c}{Non-zero elements of $15 \times 15$ matrices $\beta_\mu$ equal
to $\pm 1$.}\\
\hline
\multicolumn{2}{c}{} & \multicolumn{2}{c}{}&
\multicolumn{2}{c}{}& \multicolumn{2}{c}{}& \multicolumn{2}{c}{}\\[-2mm]
\multicolumn{2}{c}{$\beta_1$} & \multicolumn{2}{c}{$\beta_2$} &
\multicolumn{2}{c}{$\beta_3$} &
\multicolumn{2}{c}{$\beta_4$} & \multicolumn{2}{c}{$\beta_5$}\\[1mm]
\hline
&&&&&&&&&\\[-3mm]
$\!\!\!\!\phantom{1}4,15$   & $15,4\phantom{0}$  & $\phantom{-1}3,15$   &$\phantom{-}15,3\phantom{0}$   & $\phantom{-}2,15$ &$\phantom{-1}5,12$         & $\phantom{-}1,15$ &$\phantom{-}15,1$  &$-1,14$    &$-14,1\!\!\!\!$\\
$\!\!\!\!\phantom{1}7,14$   & $14,7\phantom{0}$ & $\phantom{-1}6,14$    &$\phantom{-}14,6\phantom{0}$   &$\;15,14$      &$\phantom{-}14,55$         & $-5,13$       &$-13,5$        &$-2,13$        & $-13,2\!\!\!\!$\\
$\!\!\!\!\phantom{1}9,13$   & $13,9\phantom{1}$  & $\phantom{-1}8,13$ &$\phantom{-}13,8\phantom{0}$ &$-8,12$        &$-12,8\phantom{2}$             & $-6,12$       & $-12,6$       & $-3,12$       & $-12,3\!\!\!\!$\\
$\!\!\!\!10,12$         & $12,10$       & $-10,11$      &$-11,10$               & $-9,11$       & $-11,9\phantom{2}$            & $-7,11$       & $-11,7$       &$-4,11$    & $-11,4\!\!\!\!$
\end{tabular}
\end{center}

One can clearly see from the explicit form of diagonal matrices ${\pbf S}^2$, ${\pbf T}^2$,
$S_3$, $T_3$, $\varepsilon =\beta_5$ that eq.~(34) with $15\times 15$ matrices
$\beta_\mu$ realize the representation
\be
D^+\left( \frac 12 , \frac 12 \right)\oplus D^- \left( \frac 12 , \frac 12 \right)\oplus
D(1,0) \oplus D(0,1) \oplus D(0,0),
\ee
where the first two representations of the group $P(1,4)$ are realized by eight principal
components of the vector function $\Phi$ and the last three representations of the group
$O(4)$ are realized by seven redundant components of the vector function $\Phi$.
 Of course, only the eight components realizing the representation
\be
D^+\left( \frac 12 , \frac 12 \right)\oplus D^- \left( \frac 12 , \frac 12 \right)
\ee
have a physical sense. From the seven redundant components the eight principal can
be separated by transformation of the Foldy--Wouthuysen type
\be
U=\exp \left( -i \frac{\beta_k p_k}{2p}\; \mbox{arctg}\; \frac{p}{\varkappa}\right),
\qquad p\equiv \sqrt{p_k^2}.
\ee
This transformation splits in an invariant way eq.~(34) with $15\times 15$
matrices into two independent equations, the first being for the principle components
$\psi(x_0,{\pbf x}, x_4,s_3, t_3)$, $s_3,t_3=\pm \frac 12$  and coincide with eq.~(8) in ref.~[1],
and the second being for redundant components
 $\phi(x_0,{\pbf x}, x_4,s_3, t_3)$, $s_3,t_3=0, \pm 1$, and
$\phi_0(x_0,{\pbf x}, x_4)$ having no physical sense.

Thus, the Kemmer--Duffin equation (34) in Minkowski five-space with $15\times 15$
matrices describes symmetrically fermions and antifermions with spin and isospin $s=t=\frac 12$
(multiplets of the type spinodoublet-isodoublet), i.e., for example, the systems of particles
like a nucleon-antinucleon $(N,\overline N)$. This equation is, of course, $PTC$ invariant.

As it was mentioned above, in five-space the algebra KDP eq.~(35) has three irreducible
representations. The third irreducible representation is realized by $20\times 20$ matrices
$\beta_\mu$. We do not present here the explicit form of the matrices and the analysis
of the equation connected with them. Not only that the principal components of the wave
function of this equation realize the representation
\be
D^+(1,0)\oplus D^-(1,0) \oplus D^+(0,1) \oplus D^-(0,1)
\ee
of the group $P(1,4)$, i.e.,  describe a meson multiplet like $(\pi, \omega)$, and is
$PTC$ invariant as well.

In this paper we have made the analysis of $P(1,4)$-invariant equations of the Dirac and
Kemmer--Duffin type. The analysis of another linear on $\p_\mu$ equations in five-dimensional
space, for example, equations of the Rarita--Schwinger type, Pauli--Fierz type and other,
can be made analogously. It is interesting to note that the Rarita--Schwinger formalism developed
in the $P(1,3)$ scheme for finding equations for particles with arbitrary spin, can be
generalized on the case of the $P(1,4)$ scheme without any difficulties. This is because of
there are five matrices $\gamma_\mu$, $\mu=0,1,2,3,4$, obeying the algebra ~(16),
and in the Rarita--Schwinger formalism for the  $P(1,4)$
scheme all the five matrices are equal in rights. Note that in the case of the KDP algebra~(35)
the situation is another: there is no fifth $5\times 5$ or $10\times 10$ matrix $\beta_5$
obeying the algebra~(35).

It should finally be noted that the general form of the $P(1, n)$-invariant equation linear in
$\p_\mu$ is
\be
(B_\mu \p_\mu +\varkappa)\Phi=0, \qquad \mu=1,2,\ldots, n,n+1,
\ee
where the operators $B_\mu$ are defined by the relations
\be
[B_\mu, S_{\rho\sigma}]_-=\delta_{\mu\rho} B_\sigma -\delta_{\mu\sigma}B_\rho,
\qquad \mu,\rho,\sigma=1,\ldots, n+1.
\ee
For the representation of class I the operators $B_\mu$ are finite-dimensional, for those of
class~III the operators $B_\mu$ can be both finite- and infinite-dimensional. Definite forms
of operators $B_\mu$ can be found by the method proposed in ref.~[15].

For the case of the group $P(1,4)$ eqs.~(44) referred to either $B_\mu$ or $S_{\mu\nu}$
describe particles with either values of spin $s$ or isospin $t$. These equations, however,
contain a lot of redundant components. The analysis of eqs.~(44) with matrices $B_\mu$
answering the question which representation of the group $P(1,4)$ is realized by the equation,
and the $P(1,4)$-invariant split of the equation in principal and redundant parts, can be
made with the help of the method illustrated here for the case of the Kemmer-Duffin
equations~(34).

As was shown in ref. [16], eq.~(44) or any equation on the group $P(1,4)$
is invariant under the discrete operators $P$, $T$, $C$ if $\Phi$ transforms by the following
representation of the group $P(1,4)$
\be
D^+(s,t)\oplus D^-(s,t) \oplus D^+(t,s) \oplus D^-(t,s).
\ee

\begin{enumerate}
\footnotesize

\item Fushchych W.I., Krivsky I.Yu., {\it Nucl. Phys. B}, 1968, {\bf 7},
79.\ \ {\tt quant-ph/0206057}

\item Wigner E.P., {\it Ann. Math.}, 1939, {\bf 40}, 149.

\item Shirokov Yu.M., {\it JETP (Sov. Phys.)}, 1957, {\bf  33}, 1196.

\item Joos H., {\it Fortschr. Phys.}, 1962, {\bf 10}, 65.

\item Foldy L., {\it Phys. Rev.}, 1956, {\bf  102}, 568.

\item Gelfand I.M., Zeitlin M., {\it DAN USSR}, 1950, {\bf  71}, 1017.

\item Gelfand I.M., Grayev M.I., {\it Trudy Moskovskogo Matematicheskogo obshchestva},
1959, {\bf  8}; {\it Izv. AN USSR, Ser. Math.}, 1965, {\bf  29},  5.

\item Dirac P.A.M., {\it Proc. Roy. Soc. A},  1936, {\bf 155}, 447.

\item Gelfand I.M., Minlos R.A., Shapiro Z.Ya., Representations of the rotation and Lorentz
group and their applications, Moscow, Fizmatgiz, 1958 (in Russian).

\item Umezawa H., Quantum field theory, North-Holland, Amsterdam, 1956.

\item Fushchych W.I., Preprint ITF 69-17, Kiev, 1969.\ \ {\tt quant-ph/0206045}

\item d'Espagnat B., Prentki J., {\it Phys. Rev.}, 1955, {\bf  99}, 328.

\item Foldy L., Wouthuysen S., {\it Phys. Rev.}, 1950, {\bf 78}, 29.

\item Case K.M., {\it Phys. Rev.}, 1955, {\bf 100}, 1513.

\item Fushchych W.I., {\it Ukrain. Phys. J.}, 1966, {\bf  8}, 907.

\item Fushchych W.I., {\it J. Theor. Math. Phys.}, 1970, {\bf 4}, 360

\end{enumerate}

\end{document}